\magnification=\magstep1
\input amstex
\documentstyle{amsppt}
\NoBlackBoxes
\pageheight {8.5 truein}
\pagewidth {6.0 truein}

\topmatter
\title
Noncommutative Toda chains, Hankel quasideterminants and Painlev\'e II equation
\endtitle
\author
Vladimir Retakh and Vladimir Rubtsov
\endauthor
\leftheadtext{Vladimir Retakh and Vladimir Rubtsov}
\rightheadtext{Noncommutative Painlev\'e equation}
\address
\newline
V. Retakh: Department of Mathematics, Rutgers University,
Piscataway,
NJ 08854-8019
USA
\newline
V. Rubtsov: D\'epartement de Mat\'ematiques, Universit\'e d'Angers,
LAREMA UMR 6093 du CNRS,
2, bd. Lavoisier, 49045, Angers, Cedex 01,
France and Theory Division, ITEP, 25, B. Tcheremushkinskaya, 117259, Moscow, Russia
\endaddress
\email
\newline vretakh\@math.rutgers.edu
\newline
Volodya.Roubtsov\@univ-angers.fr
\endemail
\abstract
We construct solutions of an infinite Toda system and an analogue of the Painlev\'e
II equation over noncommutative differential division rings in terms of
quasideterminants of Hankel matrices.
\endabstract

\endtopmatter

\document
\head Introduction\endhead

Let $R$ be an associative algebra over a field with a derivation
$D$. Set $Df=f'$ for any $f\in R$. Assume that $R$ is a division
ring. In this paper we construct solutions for the system of
equations (0.1) over algebra $R$
$$
(\theta _n'\theta _n^{-1})'=\theta _{n+1}\theta _n^{-1}-\theta _n\theta _{n-1}^{-1},\ \ \  n\geq 1\tag 0.1-n
$$
assuming that $\theta_1=\phi, \theta_0=\psi^{-1}$, $\phi,\psi \in R$
and its ``negative" counterpart (0.1')
$$(\eta _{-m}^{-1}\eta _{-m}')'=\eta _{-m}^{-1}\eta _{-m-1}
-\eta _{-m+1}^{-1}\eta _{-m}, \ \ m\geq 1 \tag 0.1'-m$$
where $\eta_0=\phi^{-1}, \eta_{-1}=\psi$.

Note that $\theta '\theta ^{-1}$ and $\theta ^{-1}\theta '$ are noncommutative analogues of the logarithmic derivative
$(\log \theta)'$.

We use then the solutions of the Toda equations under a certain anzatz for constructing
solutions of the {\it noncommutative Painlev\'e II equation}
$$
P_{II}(u,\beta):\ \ \ u''=2u^3-2xu-2ux+4(\beta + \frac {1}{2})
$$
where $u,x\in R$, $x'=1$ and $\beta$ is a scalar parameter, $\beta '=0$.

Unlike papers \cite {NGR} and \cite{N} we consider here a ``pure noncommutative" version
of the Painlev\'e equation without any additional assumption for our algebra $R$.

In fact a noncommutative ("matrix") version of Painlev\'e II

$$P_{II}(u,\beta):\ \ \ u''=2u^3 + xu +{\beta}I.$$

was considered in the first time in the papers of V. Sokolov with different coauthors:
We mention here f.e. \cite{BS}. But their form of this equation, satisfying the Painlev\'e
test, in the same time can not be obtained as a reduction of some matrix analog of mKDV system.

Our equation is similar to this noncommutative Painlev\'e II but there is an essential
difference: we write the second term in the R.H.S. in the symmetric or "anticommutator" form.
This splitting form is much more adaptable to some generalizations of the usual commutative
Painlev\'e II.

Our motivation is the following. In the commutative case one can consider an
infinite Toda system (see, for example \cite {KMNOY, JKM}):
$$\tau _n''\tau _n-(\tau_n')^2=\tau _{n+1}\tau _{n-1}-\phi \psi \tau _n^2 \tag 0.2-n$$
with the conditions $\tau_1=\phi, \tau _0=1, \tau_{-1}=\psi$.

Let $n\geq 1$. By setting $\theta _n=\tau _n/\tau _{n-1}$ the system can be written as
$$(\log \tau _n)''=\theta _{n+1}\theta _n^{-1}-\phi \psi.$$

For $n=1$ we have the equation (0.1-1)with $\theta_1=\phi, \theta_0=\psi^{-1}$.
By subtracting equation (2.2-n) from (2.2-(n+1)) and replacing
the difference $\log \tau _{n+1}-\log \tau _n$ by $\log \frac {\tau _{n+1}}{\tau _n}$ one can get (0.1-n).

Similarly, the system (0.2-m) for positive $m$ implies the system $(0.1'-m)$ for
$\theta _{-m}=\tau _{-m}/\tau _{-m+1}$.

By going from $\tau _n$'s to their consequtive relations we are cutting the system of equations
parametrized by $-\infty <n<\infty$ to its ``positive" and ``negative"
part.

A special case of the semi-infinite system (0.1) over noncommutative algebra
with $\theta_0^{-1}$ formally equal to zero
was treated in \cite {GR2}. In this paper solutions of the Toda system (0.1) with $\theta_0^{-1}=0$
were constructed as {\it quasideterminants} of certain Hankel matrices. It was the first
application of quasideterminants introduced in \cite {GR1} to noncommutative integrable systems.
This line was continued by several reseachers, see, for example, \cite {EGR1, EGR2}, papers
by Glasgow school \cite {GN, GNO, GNS} and a recent \cite{DFK}.

In this paper we generalize the result of \cite {GR2} for $\theta _0=\psi ^{-1}$ and extend
it to the infinite Toda system. The solutions are also given in terms of quasideterminants of
Hankel matrices but the computations are much harder. We follow here the commutative approach
developed in \cite {KMNOY, JKM} with some adjustments but our proofs are far from a straightforward
generalization. In particular, for our proof we have to introduce and investigate
{\it almost Hankel matrices} (see Section 2.2).

From solutions of the systems (0.1) and (0.1') under certain anzatz we deduce solutions for
the noncommutative equation $P_{II}(u,\beta)$ for various parameters $\beta$ (Theorem 3.2).
This is a
noncommutative development of an idea from \cite {KM}.

We start this paper by a reminder of basic properties of quasideterminants, then construct
solutions of the systems (0.1) and (0.1'), then apply our results to noncommutative
Painlev\'e II equations following
the approach by \cite {KM}.

Our paper shows that a theory of ``pure" noncommutative Painlev\'e
equations and the related tau-functions can be rather rich and
interesting. The Painlev\'e II type was chosen as a model and we are going
to investigate other types of Painlev\'e equations.

\noindent
{\bf Acknowledgements}. Both authors are thankful to MATPYL project
``Noncommmutative Integrable Systems" (2008-2010) for a
support of visits of V. Retakh to Angers. We are grateful
to the Federation of Mathematics of Loire Region and to LAREMA
for the help and warm hospitality.
V. Rubtsov was partially supported during the period of this work
by PICS Project ``Probl\`emes de Physique Math\'ematiques" (France-Ukraine).
He enjoyed in 2009-2010 a CNRS delegation at LPTM of Cergy-Pontoise
University and he acknowledges  a warm hospitality
and stimulating atmosphere of LPTM. He is thankful to M. Kontsevich and
to S. Duzhin who inspired his interest in noncommutative integrable systems
and to B. Enriquez for a long collaboration and discussions which triggered
him to study application of quasideterminants in quantum integrability.
Vladimir Retakh would like to thank IHES for its hospitality during his
visit in 2010.

We thank the referee for the careful reading of the manuscript and helpful remarks.

\head 1. Quasideterminants \endhead

The notion of quasideterminants was introduced in \cite {GR1}, see also
\cite {GR2-3, GGRW}.

Let $A=||a_{ij}||$, $i,j=1,2,\dots , n$ be a matrix over an associative
unital ring. Denote by $A^{pq}$ the $(n-1)\times (n-1)$ submatrix of
$A$ obtained by deleting the $p$-th row and $q$-th column. Let $r_i$ be
the row matrix $(a_{i1}, a_{i2}, \dots \hat a_{ij},\dots, a_{in})$ and
$c_j$ be the column matrix with entries
$(a_{1j}, a_{2j}, \dots \hat a_{ij},\dots, a_{nj})$.

For $n=1$, $|A|_{11}=a_{11}$. For $n>1$ the {\it quasideterminant} $|A|_{ij}$ is defined if the matrix $A^{ij}$
is invertible. In this case
$$|A|_{ij}=a_{ij}-r_i(A^{ij})^{-1}c_j.$$

If the inverse matrix $A^{-1}=||b_{pq}||$ exists then
$b_{pq}=|A|_{qp}^{-1}$ provided that the quasideterminant is invertible.

If $R$ is commutative then $|A|_{ij}=(-1)^{i+j}\det A/\det A^{ij}$ for any $i$ and
$j$.

\example{Examples} (a) For the generic $2\times 2$-matrix $A=(a_{ij})$,
$i,j=1,2$, there are four quasideterminants:
$$
\matrix
|A|_{11} = a_{11} - a_{12}a_{22}^{-1}a_{21},\quad
&|A|_{12}=a_{12}-a_{11}a_{21}^{-1}a_{22},\\
|A|_{21} = a_{21} - a_{22}a_{12}^{-1} a_{11},\quad &
|A|_{22}=a_{22}-a_{21}a_{11}^{-1}a_{12}.
\endmatrix
$$

(b) For the generic $3\times 3$-matrix $A=(a_{ij})$, $i,j=1,2,3$, there
are 9 quasideterminants. One of them is
$$
\align
|A|_{11}=a_{11}&-a_{12}(a_{22}-a_{23}a_{33}^{-1}a_{32})^{-1}a_{21}
-a_{12}(a_{32}-a_{33}a_{23}^{-1} a_{22})^{-1} a_{31}
\\
& -a_{13}(a_{23}-a_{22}a_{32}^{-1}a_{33})^{-1}a_{21}
-a_{13}(a_{33}-a_{32}\cdot a_{22}^{-1}a_{23})^{-1}a_{31}.
\endalign
$$
\endexample

Here are the transformation properties of quasideterminants.
Let $A=||a_{ij}||$ be a square matrix of order $n$ over a ring $R$.

(i)  The quasideterminant $|A|_{pq}$ does not depend on permutations
of rows and columns in the matrix $A$ that do not involve the $p$-th
row and the $q$-th column.

(ii)  {\it The multiplication of rows and columns.}  Let
the matrix $B=||b_{ij}||$ be obtained from the matrix $A$ by
multiplying the $i$-th row by $\lambda\in R$  from
the left, i.e., $b_{ij}=\lambda a_{ij}$ and $b_{kj}=a_{kj}$ for
$k\neq i$. Then
$$
 |B|_{kj}=\cases \lambda |A|_{ij} \quad&\text{ if } k = i,\\
|A|_{kj} \quad&\text{ if } k \neq i \text{ and } \lambda \text{ is
invertible.}\endcases
$$

Let the matrix $C=||c_{ij}||$ be obtained from the matrix A by
multiplying the $j$-th column by $\mu\in R$ from the
right, i.e. $c_{ij}=a_{ij}\mu $ and $c_{il}=a_{il}$ for all $i$
and $l\neq j$. Then
$$
|C|_{i\ell}=\cases |A|_{ij} \mu \quad&\text{ if } l = j,\\
|A|_{i\ell} \quad&\text{ if } l \neq j \text{ and } \mu \text{ is
invertible.}\endcases
$$

(iii) {\it The addition of rows and columns.} Let the matrix
$B$ be obtained from $A$ by replacing the $k$-th row of $A$ with the
sum of the $k$-th and $l$-th rows, i.e., $b_{kj}=a_{kj}+a_{lj}$,
$b_{ij}=a_{ij}$ for $i\neq k$. Then
$$
|A|_{ij} = |B|_{ij}, \qquad  i=1, \dots k-1, k+1,\dots n,
\quad j=1, \dots, n.
$$

We will need the following property of quasideterminants sometimes called
the {\it noncommutative Lewis Carroll identity}. It is a special case of
the {\it noncommutative Sylvester identity} from \cite {GR1-2} or {\it
heredity principle} formulated in \cite {GR3}.

Let $A=||a_{ij}||$,
$i,j=1,2,\dots ,n$. Consider the followng $(n-1)\times (n-1)$-submatrices
$X=||x_{pq}||$, $p,q=1,2,\dots , n-1$ of $A$: matrix $A_0=||a_{pq}||$ obtained from
$A$ by deleting its $n$-th row and $n$-th column;
matrix $B=||b_{pq}||$ obtained from $A$ by
deleting its $(n-1)$-th row and $n$-th column;
matrix $C=||c_{pq}||$ obtained from $A$ by
deleting its $n$-th row and $(n-1)$-th column;
matrix $D=||d_{pq}||$ obtained from $A$ by
deleting its $(n-1)$-th row and $(n-1)$-th column. Then
$$
|A|_{nn}=|D|_{n-1, n-1}-|B|_{n-1,n-1}|A_0|_{n-1,n-1}^{-1}|C|_{n-1,n-1} \tag 1.1.
$$

\head 2. Quasideterminant solutions of noncommutative Toda equations\endhead

\subhead 2.1. Noncommutative Toda equations in bilinear form \endsubhead
Let $F$ be a commutative field and $R$ be an associative ring containing $F$-algebra.
Let $D:R\rightarrow R$ be a derivation over $F$,
i.e. an $F$-linear map satisfying the Leibniz rule $D(ab)=D(a)\cdot b+a\cdot D(b)$ for any $a,b\in R$. Also,
$D(\alpha)=0$ for
any $\alpha \in F$. As usual, we set $u'=D(u), u''=D(D(u)), \dots $. Recall that $D(v^{-1})=-v^{-1}v'v^{-1}$ for any
invertible $v\in R$.

Let $\phi, \psi \in R$ and $R$ be a division ring. We construct now solutions for the noncommutative Toda
equations (0.1) and (0.1')
assuming that $\theta _0=\psi ^{-1}$, $\theta _1=\phi$ and
$\eta _0=\phi^{-1}$, $\eta_{-1}=\psi$.

Set (cf. \cite {KMNOY, JKM} for the commutative case) $a_0=\phi, b_0=\psi$ and
$$a_n=a_{n-1}'+\sum _{i+j=n-2, i,j\geq 0}a_i\psi a_j, \ \
      b_n=b_{n-1}'+\sum _{i+j=n-2, i,j\geq 0}b_i\phi b_j, \ \ n\geq 1. \tag 2.1 $$

Construct Hankel matrices $A_n=||a_{i+j}||$, $B_n=||b_{i+j}||, i,j=0,1,2\dots , n$.

\proclaim{Theorem 2.1}
Set $\theta _{p+1}=|A_p|_{p,p}$, $\eta _{-q-1}=|B_q|_{q, q}$. The elements $\theta _n$ for $n\geq 1$
satisfy the system (0.1) and the elements $\eta _{-m}, m\geq 1$ satisfy the system (0.1').
\endproclaim

This theorem can be viewed as a noncommutative generalization of Theorem 2.1 from \cite {KMNOY}.
In \cite {KMNOY} it was proved that in the commutative case the Hankel determinants $\tau _{n+1}=\det A_n, n\geq 0, \tau _0=1,
\tau _{-n-1}=\det B_n$, $n\leq 0$ satisfy the system (0.2).

\example{Example}
The (noncommutative) logarithmic derivative $\theta_1'\theta_1^{-1}$ satisfies the noncommutative Toda equation (0.1-1):
$$(\theta_1'\theta_1^{-1})' = \theta_2\theta_1^{-1} - \phi\psi.$$
\endexample

In fact,
$$
(\theta_1'\theta_1^{-1})' = (a_1 a_0^{-1})'= (a_2 - a_0\psi a_0)a_0^{-1} - (a_1 a_0^{-1})^2
$$
$$
=(a_2 - a_1 a_0^{-1}a_1)a_0^{-1} - a_0\psi = \theta_2\theta_1^{-1} - \phi\psi.
$$

Our proof of Theorem 2.1. in the general case is based on properties of quasideterminants of almost Hankel matrices.

\subhead 2.2. Almost Hankel matrices and their quasideterminants\endsubhead
We define {\it almost Hankel} matrices $H_n(i,j)=||a_{st}||$,
$s,t=0,1,\dots , n$, $i,j\geq 0$ for a sequence $a_0, a_1, a_2, \dots $
as follows. Set $a_{nn}=a_{i+j}$ and for $s,t<n$

$$a_{s,t}=a_{s+t}, \ \ a_{n,t}=a_{i+t}, \ \ a_{s,n}=a_{s+j}.$$
and $a_{nn}=a_{i+j}$.

Note that $H_n(n,n)$ is a Hankel matrix.

Denote by $h_n(i,j)$ the quasideterminant $|H_n(i,j)|_{nn}$. Then
$h_n(i,j)=0$ if at least one of the inequalities $i<n$, $j<n$ holds.

\proclaim{Lemma 2.2}
$$h_n(i,j)'=\kappa_n(i,j) -
\sum _{p=1}^ia_{p-1}\psi h_n(i-p,j)-
\sum _{q=1}^jh_n(i,j-q)\psi a_{q-1} \tag 2.2$$
where
$$\kappa_n(i,j)=
h_n(i+1,j)-h_{n-1}(i,n-1)h_{n-1}^{-1}(n-1,n-1)h_n(n,j).\tag 2.3a$$
Also,
$$=
h_n(i,j+1)-h_n(i,n)h_{n-1}^{-1}(n-1,n-1)h_{n-1}(n-1,j).\tag 2.3b
$$
\endproclaim

Note that some summands $h_n(i-p,j)$, $h_n(i,j-q)$ in formula (2.2)
can be equal to zero.

Since $h_n(i,j)=0$ when $i<n$ or $j<n$ we have the following corollary.

\proclaim{Corollary 2.3} $$h_n(n,n)'=\kappa _n(n,n),$$
$$
h_n(i,n)'=\kappa _n(i,n)-\sum _{s=1}^ia_{s-1}\psi h_n(i-s,n),
$$
$$
h_n(n,j)'=\kappa _n(n,j)-\sum _{v=1}^jh_n(n,j-v)\psi a_{v-1}.
$$
\endproclaim

\demo {Proof of Lemma 2.2} We prove Lemma 2.2 by induction. By definition,
$$h_1(i,j)'=
a_{i+j+1}-\sum _{k=0}^{i+j-1}a_k\psi a_{i+j-1-k}-
(a_{i+1}-\sum _{s=0}^{i-1}a_s\psi a_{i-1-s})a_0^{-1}a_j
$$
$$
+a_ia_0^{-1}a_1a_0^{-1}a_j
-a_ia_0^{-1}(a_{j+1}-\sum _{t=0}^{j-1}a_{j-1-t}\psi a_t).
$$

Set
$$\kappa _1(i,j)=a_{i+j+1}-a_{i+1}a_0^{-1}a_j+a_ia_0^{-1}a_1a_0^{-1}a_j
-a_ia_0^{-1}a_{j+1},
$$
we can check formulas (2.3a) and (2.3b). The rest of the proof for $n=1$ is easy.

Assume now that formula (2.2) is true for $n\geq 1$ and prove it for $n+1$.
By the noncommutative Sylvester identity (1.1)
$$h_{n+1}(i,j)=h_n(i,j)-h_n(i,n)h_n^{-1}(n,n)h_n(n,j). \tag 2.4 $$

Set $h_{n+1}(i,j)'=\kappa _{n+1}(i,j)+r_{n+1}(i,j)$ where $\kappa _{n+1}$
contains all terms without $\psi$. Then
$$\kappa _{n+1}(i,j)=\kappa _n(i,j)-\kappa _n(i,n)h_n^{-1}(n,n)h_n(n,j)$$
$$+h_n(i,n)h_n^{-1}(n,n)\kappa _n(n,n)h_n^{-1}(n,n)h_n(n,j)
-h_n(i,n)h_n^{-1}(n,n)\kappa _n(n,j).
$$

By induction, the first two terms can be written as
$$
h_n(i+1,j)-h_{n-1}(i,n-1)h_{n-1}^{-1}(n-1,n-1)h_n(n,j)
$$
$$+[h_n(i+1,n)-h_{n-1}(i,n-1)h_{n-1}^{-1}(n-1,n-1)h_n(n,n)]
h_n^{-1}(n,n)h_n(n,j)
$$
$$
=h_n(i+1,j)-h_n(i+1,n)h_n^{-1}(n,n)h_n(n,j).
$$
This expression equals to $h_{n+1}(i+1,j)$ by the Sylvester identity.

The last two terms in $\kappa _{n+1}(i,j)$ can be written as
$$
h_n(i,n)h_n^{-1}(n,n)[h_n(n+1,n)-h_{n-1}(n,n-1)h_{n-1}^{-1}(n-1,n-1)h_n(n,n)]
h_n^{-1}(n,n)h_n(n,j)
$$
$$
-h_n(i,n)h_n^{-1}(n,n)[h_n(n+1,j)-h_{n-1}(n,n-1)h_{n-1}^{-1}(n-1,n-1)h_n(n,j)]
$$
$$
=h_n(i,n)h_n^{-1}(n,n)[-h_n(n+1,n)+h_n(i,n)h_n^{-1}(n,n)h_n(n+1,j)]
$$
$$
=-h_n(i,n)h_n^{-1}(n,n)h_{n+1}(n+1,j)
$$
also by the Sylvester identity.

Therefore,  $\kappa_{n+1}(i,j)$ satisfies formula (2.3a). Formula (2.3b) can be obtained in a similar way.

Let us look at the terms containing $\psi$. According to the inductive assumption
$$
h_n(i,j)'=\kappa _n(i,j)-\sum _{k=1}^ia_{k-1}\psi h_n(i-k,j)-\sum _{\ell =1}^jh_n(i,j-\ell)\psi a_{\ell -1}.
$$
Using the Corollary 2.3 and formula (2.2) for $n$ one can write $r_{n+1}(i,j)$ as
$$
-\sum _{k=1}^ia_{k-1}\psi h_n(i-k,j)-\sum _{\ell =1}^jh_n(i,j-\ell)\psi a_{\ell -1}
$$
$$
+\sum _{k=1}^ia_{k-1}\psi h_n(i-k,n)]h_n^{-1}(n,n)h_n(n,j)
$$
$$
+h_n(i,n)h_n^{-1}(n,n)\sum _{\ell=1}^j h_n(n, j-\ell)\psi a_{\ell-1}
$$
$$
=-\sum _{k=1}^ia_{k-1}\psi [h_n(i-k,j)-h_n(i-k,n)h_n^{-1}(n,n)h_n(n,j)]
$$
$$
-\sum _{\ell=1}^j[h_n(i,j-\ell)-h_n(i,n)h_n^{-1}(n,n)h_n(n,j-\ell)]\psi a_{\ell-1}.
$$

Our lemma follows now from the Sylvester identity applied to each expression in square brackets.
\enddemo

Corollary 2.3 and formula (2.3a) immediately imply

\proclaim{Corollary 2.4} For $n>1$
$$
h_n(n,n)'h_n^{-1}(n,n)=h_n(n+1,n)h_n^{-1}(n,n)-h_{n-1}(n,n-1)h_{n-1}^{-1}(n-1,n-1).
$$
\endproclaim

Note in the right hand side we have a difference of left quasi-Pl\"ucker coordinates
(see \cite {GR3}).

\subhead 2.3. Proof of Theorem 2.1\endsubhead

Our solution of the Toda system (0.1) follows from Corollary 2.4 and the following lemma.

\proclaim{Lemma 2.5} For $k>0$
$$
[h_k(k+1,k)h_k^{-1}(k,k)]'=h_{k+1}(k+1,k+1)h_k^{-1}(k,k) - a_0\psi.
$$
\endproclaim

\demo {Proof} Corollary 2.3 and formula (2.3b) imply
$$
h_k(k+1,k)'=h_k(k+1,k+1)-h_k(k+1,k)h_{k-1}^{-1}(k-1,k-1)h_{k-1}(k-1,k)-a_0\psi h_k(k,k)
$$
because $h_k(k+1-s,k)=0$ for $s>1$.

Then, using again formula (2.3b) one has
$$
[h_k(k+1,k)'h_k^{-1}(k,k)]'=
$$
$$
[h_k(k+1,k+1)-h_k(k+1,k)h_{k-1}^{-1}(k-1,k-1)h_{k-1}(k-1,k) - a_0\psi h_k(k,k)]h_k^{-1}(k,k)
$$
$$
-h_k(k+1,k)h_k^{-1}(k,k)][h_k(k,k+1)-h_k(k,k)h_{k-1}^{-1}(k-1,k-1)h_{k-1}(k-1,k)]
h_k^{-1}(k,k)
$$
$$
=[h_k(k+1,k+1)-h_k(k+1,k)h_k^{-1}(k,k)h_k(k,k+1)]h_k^{-1}(k,k)-a_0\psi=
$$
$$
h_{k+1}(k+1,k+1)h_k^{-1}(k,k)-a_0\psi
$$
by the Sylvester formula.
\enddemo

Theorem 2.1. now follows from Corollary 2.4 and Lemma 2.5.
The statement for $\eta _{-m}$, $m\geq 1$ can be proved in a similar way.

\head 3. Noncommutative Painlev\`e II\endhead

\subhead 3.1 Commutative Painlev\`e II and Hankel determinants: motivation\endsubhead
The Painlev\`e II ($P_{II}$) equation (with commutative variables)
$$u'' = 2u^3 - 4xu + 4(\beta + \frac{1}{2})$$ admits unique rational solution for
a half-integer value of the parameter $\beta$. These solutions can be expressed in terms of
logarithmic derivatives of ratios of Hankel-type determinants. Namely, if $\beta = N+\frac{1}{2}$ then
$$u =\frac{d}{dx}\log\frac{\det A_{N+1}(x)}{\det A_N(x)},$$
where $A_N (x) = ||a_{i+j}||$ where $i,j=0,1,\dots , n-1$. The entries of the matrix are polynomials
$a_n (x)$ subjected to the recurrence relations:
$$a_0 = x,\quad a_1 = 1, a_n = a_{n-1}' + \sum_{i=0}^{n-2}a_i a_{n-2-i}.$$ (see \cite{JKM})

\subhead 3.2 Noncommutative and ``quantum" Painlev\`e II\endsubhead
We will consider here a {\it noncommutative} version of $P_{II}$ which
we will denote ${\text{nc}}-P_{II}(x,\beta)$:
$$ u'' = 2u^3 -2xu - 2ux + 4(\beta +\frac{1}{2}),$$
where $x,u\in R,\quad x' = 1$ and $\beta$ is a central scalar parameter ($\beta\in F, \beta'=0$).

This equation is a specialization of a general noncommutative Painlev\'e II system with respect to
three dependent noncommutative variables $u_0,u_1,u_2$:

$$\aligned
u_0' = u_0 u_2 + u_2 u_0 + \alpha_0 \\
u_1' = -u_1 u_2 - u_2 u_1 + \alpha_1 \\
u_2' = u_1 - u_0.
\endaligned$$

Indeed, taking the derivative of the third and using the first and second, we get
$$ u_2'' = -(u_0 + u_1)u_2 -u_2(u_0 + u_1) + \alpha_1 -\alpha_0.$$
Then we have:
$$(u_0 + u_1)' = -u_2'u_2 -u_2 u_2' + \alpha_0 + \alpha_1 $$
and, immediately
$$-(u_0 + u_1) = u_2^{2} - (\alpha_0 + \alpha_1)x - \gamma,\ \  \gamma\in F.$$

Compare with $u_2''$ we obtain the following  ${\text{nc}}-P_{II}:$
$$ u_2'' = 2u_2^3 -(\alpha_0 + \alpha_1)xu_2 -(\alpha_0 + \alpha_1)u_2 x - 2\gamma u_2 + \alpha_1 -\alpha_0.$$

Our equation corresponds the choice $\gamma = 0,\alpha_1 = 2(\beta +1),\alpha_0 = -2\beta.$

\proclaim{Remark} The noncommutative Painlev\'e II system above is the straightforward generalization of the
analogues system in \cite{NGR} when the variables $u_i, i= 0,1,2$ are subordinated to some commutation
relations. Here we don't assume that the ``independent" variable $x$ commutes with $u_i$.
\endproclaim

Going further with this analogy we will write a ``fully non-commutative" Hamiltonian of the system
$$
H= \frac{1}{2}(u_0 u_1 + u_1 u_0) + \alpha_1 u_2
$$
and introduce the ``canonical" variables
$$p := u_2,\, q := u_1,\,x :=\frac{1}{2}(u_0 + u_1 + u_2^2).$$
\proclaim{Proposition 3.1}
Let a triple $(x,p,q)$ be a ``solution" of the ``Hamiltonian system" with the Hamiltonian $H$ and $\alpha_1 =2(\beta +1)$.
$$\aligned
p_x = - H_q \\
q_x = H_p .
\endaligned$$

Then $p$ satisfies the ${\text{nc}}-P_{II}$:
$$p_{xx} = 2p^3 -2px -2xp + 4(\beta +\frac{1}{2}).$$
\endproclaim
\demo{Proof} Straightforward computation gives that:
$$\aligned
p_x = p^2 + 2q -2x\\
q_x = \alpha_1 -(qp + pq).\endaligned$$

Taking $p_{xx} = p_x p + pp_x + 2q_x -2$ and substituting $p_x$ and $q_x$
we obtain the result.
\enddemo

We give (for the sake of completeness) the explicit expression of the Painlev\'e Hamiltonian $H$ in the "canonical" coordinates:
$$ H(x,p,q) = qx + xq -q^2 -\frac{1}{2}(qp^2 + p^2q) + 2(\beta + 1)p. $$

\subhead 3.3 Solutions of  the noncommutative Painlev\'e and of the Toda system\endsubhead

\proclaim{Theorem 3.2}
Let $\phi$ and $\psi$ satisfy the following identities:
$$\psi^{-1}\psi'' = \phi''\phi^{-1} = 2x - 2\phi\psi, \tag 3.1$$
$$ \psi\phi' - \psi'\phi = 2\beta. \tag 3.2$$

Then for $n\in \Bbb N$

1) $u_n = \theta_n'\theta_n ^{-1}$ satisfies  ${\text nc}-P_{II}(x,\beta + n -1)$;

2) $u_{-n} = \eta_{-n}'\eta_{-n} ^{-1}$ satisfies  ${\text nc}-P_{II}(x,\beta - n)$.
\endproclaim

Let us start with the following useful (though slightly technical) lemma
\proclaim{Lemma 3.3}
Under the conditions of the Theorem 3.1 we have the chain of identities ($n\geq 0$):

1) $\theta_n'\theta_n^{-1} + \theta_{n-1}'\theta_{n-1}^{-1} = 2(\beta +n-1)\theta_{n-1}\theta_n^{-1}$

2) $\theta_n''\theta_n^{-1} = 2(x -\theta_n\theta_{n-1}^{-1})$

and also, for $n\geq 1$

3) $\eta _{-n}^{-1}\eta _{-n}'+\eta_{-n+1}\eta_{-n+1}'=-2(\beta -n+1)\eta _{-n}^{-1}\eta _{-n}^{-1}\eta _{-n+1}$

4) $\eta _{-n}^{-1}\eta _{-n}''=2(x-\eta _{-n+1}^{-1}\eta _{-n}$.
\endproclaim
\demo{Proof}
Remark that the first step in the chain ($n=1$) directly follows from our assumption: $\theta_1 =\phi,\theta_0 = \psi^{-1}:$
$$\phi'\phi^{-1} + (\psi^{-1})'\psi = 2\beta\psi^{-1}\phi^{-1}.$$
Indeed, we have
$$\phi'\phi^{-1} - \psi^{-1}\psi' = 2\beta\psi^{-1}\phi^{-1}$$
where the result:
$$\psi\phi' - \psi'\phi =2\beta.$$

The second step ($n=2$) is a little bit tricky.

We consider the Toda equation $(\phi'\phi^{-1})' = \theta_2\phi^{-1}-\phi\psi$  and find easily $\theta_2$ (using $\phi''\phi^{-1} = 2x-2\phi\psi$):
$$\theta_2 =2x\phi -\phi\psi\phi - (\phi'\phi^{-1})\phi'.$$
Taking the derivation and using the same Toda and the first step identity, we get
$$\theta_2' = 2\phi(\beta +1)- \phi'\phi^{-1}\theta_2.$$

The second ($n=2$) identity is rather straightforward:
$$\theta_2'' + (\theta_1'\theta_1^{-1})'\theta_2 + (\theta_1'\theta_1^{-1})\theta_2' =2(\beta+1)\theta_1'.$$
Again using the Toda and the first identity we obtain finally:
$$\theta_2''\theta_2^{-1} +\theta_2\phi^{-1} - \phi\psi - (\phi'\phi^{-1})^2 =0$$
and then
$$\theta_2''\theta_2^{-1} + 2x -2(\phi\psi + (\phi'\phi^{-1})^2)= \theta_2''\theta_2^{-1}-2(x-\theta_2\phi^{-1})=0.$$

We will discuss one more step, namely the passage from $n=2$ to $n=3$ (then the recurrence will be clear).
We want to show that:

1°) $\theta_3'\theta_3^{-1} + \theta_2'\theta_2^{-1} = 2(\beta + 2)\theta_2\theta_3^{-1};$

2°) $\theta_3''\theta_3^{-1} = 2(x - \theta_3\theta_2^{-1}).$

From the second Toda and second identity we get
$$\theta_3 = 2x\theta_2 -\theta_2\theta_1^{-1}\theta_2 -\theta_2'\theta_2^{-1}\theta_2'.$$
It implies
$$\theta_3' = 2\theta_2 + 2x\theta_2' -\theta_2'\theta_1^{-1}\theta_2 +
\theta_2\theta_1^{-1}\theta_1'\theta_1^{-1}\theta_2 -\theta_2\theta_1^{-1}\theta_2' -$$
$$ -2(x -\theta_2\theta_1^{-1} )\theta_2' + (\theta_2'\theta_2^{-1})^2\theta_2' - \theta_2'\theta_2^{-1}(2x - 2\theta_2\theta_1^{-1})\theta_2.$$

We simplify and obtain from this

$$
\theta _3'=2\theta _2+
\theta_2\theta _1^{-1}(\theta _2'+\theta _1'\theta _1^{-1}\theta _2)+
\theta _2'\theta_1^{-1}\theta _2+
(\theta _2'\theta _2^{-1})^2\theta _2'-2\theta _2'\theta _2^{-1}x\theta_2.
$$

By the identity for $\theta_2'$  we have
$$
\theta _3'=2\theta _2+\theta_2\theta_1^{-1}\cdot 2(1+\beta)\theta_1
+\theta_2'\theta_1^{-1}\theta_2+
\theta_2'\theta_2^{-1}(-\theta_3-\theta _2\theta_1^{-1}\theta_2).
$$
which assure the first identity for $n=3$.

Now we prove the second.

Set $a=2(\beta +2)$. We have
$$
\theta_3'=a \theta_2 - (\theta_2'\theta_2^{-1})\theta_3.
$$

Take the second derivation:
$$
\theta_3''=a\theta_2' - (\theta_2'\theta_2^{-1})'\theta_3
-\theta_2'\theta_2^{-1}\theta_3'.
$$

By using the formula for $\theta_3'$ we have
$$
\theta_3''=a\theta_2' - (\theta_2'\theta_2^{-1})'\theta_3
-\theta_2'\theta_2^{-1}(a\theta_2-\theta_2'\theta_2^{-1}\theta_3).
$$

The terms with $a$ are cancelled and we have
$$
\theta_3''= - (\theta_2'\theta_2^{-1})'\theta_3
+(\theta_2'\theta_2^{-1})^2\theta_3.
$$

Note that
$$
-(\theta_2'\theta_2^{-1})'+(\theta_2'\theta_2^{-1})^2=
\theta_2''\theta_2^{-1}-2(\theta_2'\theta_2^{-1})'.
$$

We already know that the first summand in the right hand side
equals $2(x-\theta_2\theta_1^{-1})$ and by our Toda system
$$
(\theta_2'\theta_2^{-1})'=
\theta_3\theta_2^{-1}-\theta_2\theta_1^{-1}
$$
we obtain the second identity for $\theta _3$.

The $n-$th  step of the recurrence goes as follows: from $n-$th Toda  and recurrence conjecture
we have
$$\theta_{n+1} = 2x\theta_n -\theta_n\theta_{n-1}^{-1}\theta_n -\theta_n'\theta_n^{-1}\theta_n'.$$
It implies
$$\theta_{n+1}' = 2\theta_n + 2x\theta_n' -\theta_n'\theta_{n-1}^{-1}\theta_n +
\theta_n\theta_{n-1}^{-1}\theta_{n-1}'\theta_{n-1}^{-1}\theta_n -\theta_n\theta_{n-1}^{-1}\theta_n' -$$
$$ -2(x -\theta_n\theta_{n-1}^{-1} )\theta_n' + (\theta_n'\theta_n^{-1})^2\theta_n' - \theta_n'\theta_n^{-1}(2x - 2\theta_n\theta_{n-1}^{-1})\theta_n.$$

Then, after some simplifications we get
$$
\theta _{n+1}'=2\theta _n+
\theta_n\theta _{n-1}^{-1}(\theta _n'+\theta _{n-1}'\theta _{n-1}^{-1}\theta _n)+
\theta _n'\theta_{n-1}^{-1}\theta _n +
(\theta _n'\theta _n^{-1})^2\theta _n'-2\theta _n'\theta _n^{-1}x\theta_n.
$$

By the recurrent formula for $\theta_n'$,  we have
$$\theta _n'+\theta _{n-1}'\theta _{n-1}^{-1}\theta _n=2(\beta +1 -n)\theta _{n-1}$$ and
$$
\theta _{n+1}'=
2\theta _n+2(\beta + n-1)\theta _n
+\theta_n'\theta_{n-1}^{-1}\theta_n + (\theta_n'\theta_n^{-1})^2\theta_n' - 2\theta_n'\theta_n^{-1}x\theta_n =
$$
$$
= 2(\beta + n)\theta_n + \theta_n'\theta_{n-1}^{-1}\theta_n  +\theta_n'\theta_n^{-1}(\theta_n'\theta_{n-1}^{-1}\theta_n' -2x\theta_n)=
$$
$$
= 2(\beta + n)\theta_n + \theta_n'\theta_{n-1}^{-1}\theta_n  +\theta_n'\theta_n^{-1}(-\theta_{n+1} -\theta _n\theta_{n-1}^{-1}\theta_n) =
$$
$$
=2(\beta + n)\theta_n + \theta_n'\theta_{n-1}^{-1}\theta_n  - \theta_n'\theta_n^{-1}\theta_{n+1} - \theta_n'\theta_{n-1}^{-1}\theta_n.
$$
which assure the first identity for $n+1$.

We leave the proof of the second identity for any $n$ as an easy (though a bit lengthy) exercise similar to the case $n=3$ above.

The identities 3) and 4) can be proved in a similar way.
\enddemo

\proclaim{Lemma 3.4}
For $n=1$ the left logarithmic derivative $\phi'\phi^{-1}=: u_1$ satisfies to $nc-P_{II}(x,\beta)$.
\endproclaim
\demo{Proof} From the previous lemma we have from the first Toda equation:
$$(\phi'\phi^{-1})' = \theta_2\phi^{-1}-\phi\psi = \phi''\phi^{-1} - (\phi'\phi^{-1})^2 = 2(x-\phi\psi) - u_1^2$$
and hence
$$\theta_2\phi^{-1} = 2x -\phi\psi -u_1^2.$$

In other hand, taking the derivative of the first Toda, we get
$$u_1'' = (\theta_2\phi^{-1} -\phi\psi)' = \theta_2'\phi^{-1} -\theta_2\phi^{-1}u_1 -(\phi'\psi +\phi\psi').$$

We replace $\theta_2'\phi^{-1}$ by
$$2(\beta +1) - u_1\theta_2\phi^{-1} = 2(\beta +1) -u_1(2x -\phi\psi -u_1^2).$$
Finally we obtain
$$u_1'' = 2u_1^3 -2u_1 x - 2xu_1 + 2(\beta +1) + u_1\phi\psi + \phi\psi u_1 -(\phi'\psi +\phi\psi'),$$
but
$$u_1\phi\psi + \phi\psi u_1 -(\phi'\psi +\phi\psi') = \phi\psi\phi'\phi^{-1} -\phi\psi' = 2\beta$$
which gives the desired result.
\enddemo

Our proof of Theorem 3.2 in the general case almost {\it verbatim} repeats the proof of the Lemma 3.4.

\demo{Proof of Theorem 3.2}
Let $u_n :=\theta_n'\theta_n^{-1}.$
Now the same arguments, from the Lemma 3.4, show that:

a) $\theta_{n+1}\theta_n^{-1} = 2x -\theta_n\theta_{n-1}^{-1} - u_n^2;$

b) $\theta_{n+1}'\theta_{n}^{-1} = 2(\beta +n) - \theta_n'\theta_n^{-1}\theta_{n+1}\theta_n^{-1};$

c) $u_n'' = 2u_n^3 -2xu_n -2u_n x + 2(\beta +n) +\theta_n\theta_{n-1}^{-1}(\theta_n'\theta_n^{-1} + \theta_{n-1}'\theta_{n-1}^{-1})$.
This implies that
$u_n''= 2u_n^3 -2xu_n -2u_n x + 4(\beta +n -\frac{1}{2}).$
\enddemo

\proclaim{Remark}
Using identities 3) and 4) from Lemma 3.3 we can prove the second statement of the Theorem 3.2.
\endproclaim

\head 4. Discussion and perspectives\endhead

We have developed an approach to integrability of a fully noncommutative analog of the Painlev\'e equation.
We construct solutions of this equation related to the ``fully noncommutative" Toda chain, generalizing the
results of \cite{GR2, EGR1}. This solutions admit an explicit description in terms of Hankel quasideterminants.

We consider here only the noncommutative generalization of Painlev\'e II but it is not difficult to write down
some noncommutative analogs of other Painlev\'e transcendants. It is interesting to study their
solutions, noncommutative $\tau-$ functions, etc.
We hope that our equation (like its ``commutative" prototype) is a part of a whole noncommutative Painlev\'e hierarchy which
relates (via a noncommutative Miura transform) to the noncommutative m-KdV and m-KP hierarchies (see i.e. \cite{EGR1-2},\cite{GN},\cite{GNS}).
Another interesting problem is to study a noncommutative version of isomonodromic transformations problem for our Painlev\'e equation.
The natural approach to this problem is a noncommutative generalization of generating functions, constructed in \cite{JKM}.
The noncommutative ``non-autonomous" Hamiltonian should be studied more extensively. It would be interesting to find
noncommutative analogs of Okamoto differential equations \cite{OK}
and to generalize the description of Darboux-B\"acklund transformations
for their solutions.

We shall address these and other open questions in the forthcoming papers.

\enddocument